\documentclass[doublespacing]{elsart}
\usepackage{amsmath}
\usepackage{graphicx}
\date{}

\begin{document}

\begin{frontmatter}


 \indent

\title{Ab initio prediction on ferrotoroidic olivine Li$_4$MnFeCoNiP$_4$O$_{16}$}


\author{Hong-Jian  Feng , Fa-Min Liu }

\address{Department of Physics, School of Science, Beijing
University of Aeronautics and Astronautics, Beijing 100083, China}

\begin{abstract}
 \indent First-principles calculation predict that olivine Li$_4$MnFeCoNiP$_4$O$_{16}$
 has ferrotoroidic characteristic and ferrimagnetic configuration
 with magnetic moment of
 1.56$\mu_B$ per formula unit. The ferrotoroidicity of this
 material makes it a potential candidate for magnetoelectric
 materials . Based on the orbital-resolved density of states for the
 transtion-metal ions in Li$_4$MnFeCoNiP$_4$O$_{16}$, the spin configuration for Mn$^{2+}$,Fe$^{3+}$,Co$^{2+}$, and
 Ni$^{2+}$
 is $t_{2g}^3(\uparrow)e_g^2(\uparrow)$£¬ $t_{2g}^3(\downarrow)e_g^2(\downarrow)$£¬
 $t_{2g}^1(\downarrow)t_{2g}^3(\uparrow)e_g^1(\downarrow)e_g^2(\uparrow)$
 £¬and $t_{2g}^2(\uparrow)t_{2g}^3(\downarrow)e_g^1(\uparrow)e_g^2(\downarrow)$ ,respectively.
 Density functional theory plus U (DFT+U) shows a indirect band gap of 1.25 eV in this predicted
 material, which is not simply related to the electronic
 conductivity in terms of being used as cathode material in rechargeable
 Li-ion batteries.

\end{abstract}

\begin{keyword}
Ferrotoroidic; Density functional theory;Density of states
\PACS 71.15.Mb; 71.20.-b;75.80.+q
\end{keyword}
\end{frontmatter}


 \indent
\section{ Introduction}
 \indent
Ferrotoroidic(FTO) domains have been observed in olivine LiCoPO$_4$
recently using second harmonic generation (SHG), which are
independent of the antiferromagnetic(AFM) domains \cite{1}. The
ordered arrangement of magnetic vortices is found to be the fourth
form of ferroic order, which is currently  debated whether to be
included or not\cite{2,3,4,5,6}. The well-known three forms of
ferroic are ferromagnetism, ferroelectricity, and ferroelasticity.
This fourth form of ferroic is named as ferrotoroidicity
corresponding to the former three forms ,which have been established
well and investigated extensively. A ferrotoroidic moment is
generated by a vortex of magnetic moment, which is asymmetric under
the reversal of time and space. The time- and space- asymmetric
property relate ferrotoroidic materials deeply to the multiferroic
materials, in which ferromagnetic property violates time reversal
symmetry and ferroelectric property violates spatial inversion
symmetry.  In multiferroic materials, magnetization can be rotated
by an external electric field, and polarization can be reversed by a
reversal magnetic field, which is called the magnetoelectric effect.
Therefore, FTO materials can be used as magnetoelectric materials,
in which a magnetization is induced by an electric field, and vice
versa, and have potential applications in information storage, the
emerging field of spintronic , and sensors.

In olivine LiCoPO$_4$, the Co$^{2+}$ ions are located at positions
(1/4+$\epsilon$,1/4,-$\delta$), where $\epsilon$ and $\delta$
represent the small displacements by the $mmm$ symmetry. The four
nearest Co$^{2+}$ are arranged antiferromagnetically along the $y$
axis and have two different toroidic moments due to the rotation of
spins by $\phi = 4.6\textordmasculine$ away from $y$
axis\cite{7}.The two
 toroidic moments are opposite in direction and  not equal in value,
 leading to   the ferrotoroidic porperty. Magnetism can be
improved by substitution with the transition-metal ions by another
one such as in Bi$_2$FeCrO$_6$\cite{8} and Bi$_2$FeTiO$_6$\cite{9}.
In this paper, we put the high-spin Mn$^{2+}$ and Fe$^{3+}$ in the
sites with larger toroidic radius while low-spin Co$^{2+}$ and
Ni$^{2+}$ in the sites with smaller toroidic radius, which would
bring about a large resulting toroidic moments and further an
excellent ferrotoroidicity. This predicted olivine structural
material, Li$_4$MnFeCoNiP$_4$O$_{16}$, is expected to have
significant magnetoelectric effect as multiferroic materials due to
ferrotoroidicity, and  also have potential to be used as cathode
materials for Li-ion rechargeable batteries. In order to gain much
insight in this predicted material, we have investigated the
electronic  and  ferrotoroidic properties using first-principles
calculations. The remainder of this article is organized as
follows:in section 2, we present the computational details of our
calculations. In section 3, we report the calculated results and
discussion. In section 4, the conclusions based on our calculation
are given.
\section{Computational details}
Calculations in this work have been done using the Quantum-ESPRESSO
package\cite{10}. Our density functional theory(DFT) computations
follow the method previously reported in Ref.\cite{9}.We used our
self-interaction-corrected ultrasoft pseudopotential implementation
with the Perdew Burke Ernzerhof (PBE) exchange correlation
functional, as the common local density approximation(LDA) fails to
obtain a band gap in the transition-metal oxides. Plane-wave basis
set with kinetic energy cut-off of 38 Ry was employed. Li 2s, Mn 3s,
3p and 3d electrons,  Fe 3s, 3p and 3d electrons, Co 3s, 3p and 3d
electrons, Ni 3s, 3p and 3d electrons, P 3s and 3p electrons, and O
2s and O 2p electrons have been treated as valence states. We used
up to $6\times7\times8$ grids of special k points in total energy
calculation, and $8\times9\times10$ grids of special k points were
used for density of states(DOS) calculation. The lattice constants
and the atomic positions were taken from Ref.\cite{11} and
\cite{12}. Then, fully  structural relaxations were performed by
minimization of the Hellman-Feynman forces within a convergency
threshold of 10$^{-3}$ Ry/Bohr. The transition-metal ions were
arranged differently to find the favorable structure with minimal
energy. Consequently, the stable structural configuration is
obtained  and constructed by XCrySDen\cite{13} as in Fig.1, in which
the unit cell doubled along the $b$ axis. Collinear spin
configurations have been used to construct the different AFM order.
The favorable configuration is obtained as shown in Fig. 2
accordingly. Local density approximation(LDA) or generalized
gradient approximation(GGA) within DFT always predicts the
conduction properties in transition-metal oxides, and tends to
over-delocalize electrons when the kinetic energy of electrons is
not large enough to overcome the on-site repulsion.
 Hence, a Hubbard like term U is introduced and  taken as in Ref.\cite{14} to
 treat the transiton-metal ions, which is named as DFT+U.

\section{Results and discussion}

It is known that  Li$M$PO$_4$($M$=Mn, Fe, Co, and Ni)has an
orthorhombic cell, which contains four formula units, 28 atoms: 4
lithium, 4 $M$, 4 phosphor, and 16 oxygen atoms. Phosphor ions are
tetrahedrally coordinated to oxygen, where $M$ and lithium ions
occupy the centers of distorted oxygen octahedral with space group
$Pnma$ as shown in Fig.1. The lattice parameters and atomic
positions are from Ref.\cite{11} and \cite{12}, and we put the Mn,
Fe, Co, and Ni ions on the 4$M$ sites, respectively. The stable
structure under the condition of minimal total energy is shown in
Fig. 1. The high-spin Mn$^{2+}$ and Fe$^{3+}$ occupy the sites
having large toroidic radius, which tends to generate a net toroidic
moment as we expected. In Fig. 2, atoms in one unit cell are
projected on the XY plane to make clear the relative positions of
these four type of ions. Next,the lattice parameters and atomic
positions are fully relaxed. The relaxed lattice parameters and
Wyckoff positions are shown in Table 1. The lattice parameters are
about the averaged values  of Li$M$PO$_4$($M$=Mn, Fe, Co, and Ni)
due to the mixing of different transition-metal ions.

Let's discuss the magnetic ordering of these four type of ions in
the system. The spin configurations are constructed as no spin,
ferromagnetic, and four antiferromagnetic configurations(denoted as
NM,  FM, AFM1,  AFM2, AFM3 and AFM4, respectively.). The AFM
configuration are shown in Table 2 in detail,  and the four atomic
sites to different AFM configuration are Co,Mn,Ni,and Fe  as shown
in Fig.2 anticlockwise, and the y axis is taken as the positive
direction. The differences in the total energy in the NM, AFM1,
AFM2, AFM3, and AFM4 configuration relative to its total energy in
the FM configuration are shown in Table 2. One can see that the NM
configuration has the highest total energy and becomes the most
unstable configuration. The AFM2 is the optimum magnetic
configuration  shown in Fig.2. A resulting magnetic moment of 1.56
$\mu_B$ is obtained from calculation, which is related deeply with
the real spin configuration of the four transition-metal elements,
which will be discussed in detail in the following sections.

The spin part of the toroidic moment is expressed as
\begin{equation} \mathbf{T}\propto \sum_n
\mathbf{r}_n\times \mathbf{s}_n ,\end{equation} where $\mathbf{r}_n$
is the radius vector and  $\mathbf{s}_n$ is the spin of the $n$
magnetic ions, and the center of the unit cell is taken as the
origin.   As we know, the Mn and Fe ions are expected to be in the
higher spin states as compared with Co and Ni ions. Therefore, as
long as spins rotate away from $y$ axis, from the favorable
structure shown in Fig.2, we get
\begin{equation}
\mathbf{r}_{Mn,Fe,n}\times \mathbf{s}_{Mn,Fe,n}
>\mathbf{r}_{Co,Ni,n}\times \mathbf{s}_{Co,Ni,n} .\end{equation}
Hence, a large resulting toroidic moment can be obtained in this
predicted system as compared with the case in  LiCoPO$_4$. From the
 calculation above, we consider the ferrotoroidic and
ferrimagnetic properties can be achieved simultaneously.

The DFT+U method combines the high efficiency of LDA/GGA, and an
explicit treatment of correlation with a Hubbard-type model for a
subset of states in the system. The total energy is expressed as
\begin{equation}
E_{LDA+U}[n(r)]=E_{LDA}[n(r)]+E_{Hub}[\{n_m^{I\delta}\}]-E_{dc}[\{n^{I\delta}\}],
\end{equation}
where $n(r)$ denotes the charge density and    $n^{I\delta}$ is the
transition-metal on-site occupation matrix. From the expansion in
spherical harmonics,in first order approximation, we have
\begin{equation}
E_{LDA+U}=E_{LDA}[n(r)]+E_U[\{n_{mm'}^{I\delta}\}]=E_{LDA}[\rho]+\frac{U}{2}\sum_{I,\delta}Tr[n^{I\delta}(1-n^{I\delta})]
\end{equation}
In present work, U is 3.92, 3.71, 5.05, and 5.26 for Mn, Fe, Co, and
Ni\cite{14}, respectively.

The total density of states(DOS) using DFT and  DFT+U for
Li$_4$MnFeCoNiP$_4$O$_{16}$ are shown in Fig.3, and the DOS of
LiCOPO$_4$ is also reported for comparison. From the DOS of
LiCOPO$_4$ and Li$_4$MnFeCoNiP$_4$O$_{16}$, it is clear that the
manifold states near the Fermi level in Li$_4$MnFeCoNiP$_4$O$_{16}$
are attributed to the other transition-metal elements(Mn, Fe, and
Ni) except Co.
 It can be seen that  DFT predicts a half-metal behavior, in which
there is only a complete spin polarization of electrons at the Fermi
level. The strong localization of $3d$-electrons of these
transition-metal leads to the states passing through the Fermi
enrgy, and further to the half-metal characteristic. The strong
correlation of these $3d$ electrons are well described by employing
the DFT+U, and a indirect band gap of $\sim$1.25 eV is found in the
predicted material.

In Fig.4 we present the DFT+U band structure of
Li$_4$MnFeCoNiP$_4$O$_{16}$for majority spin. It can be seen that it
is a semiconductor with an indirect band gap $\sim$ 1.25 eV. The
valence band top and conduction band bottom lie, respectively, at S
and $\Gamma$ points of the first Brillouin zone. This suggests that
the electronic conduction performance of this new material is not
very good. A large band gap will lead to a very small number of
intrinsically generated electrons or holes when it is used as the
cathode material for Li-ion rechargeable batteries. Therefore, the
band gap will not play any significant role in setting the
concentration of conduction electrons or holes. It is likely that
the key electrons involved in transport in the predicted material
are not delocalized electrons, but localized small
polarons\cite{15}, and polaron mobility is determined by the hopping
rate of the polarons.  On the other hand,  an excellent
magnetoelectric effect can be achieved by the insulating property ,
in which the current leakage is well inhibited. Moreover, from the
band dispersion, the valence bands are very dense with little
dispersion, while the conduction bands have distinctive dispersions,
and these dispersive conduction bands are well above the Fermi
energy.

In order to gain insight in the spin configuration and the oxidation
states of the transition-metal elements in the new predicted system,
we report the  orbital-resolved DOS for Mn, Fe,  Co, and Ni in Fig.
5, Fig. 6, Fig. 7, and Fig.8, respectively. From Fig.5 , the
majority-spin(up-spin) Mn-$3d$ states in $d_{xy}, d_{yz}, d_{z^2},
d_{xz}$, and $d_{x^2-y^2}$ orbitals are all occupied, and
minority-spin(down-spin) in corresponding orbitals are all empty.
This result is the formal high-spin(S=5/2) state of Mn$^{2+}$ with
$t_{2g}^3(\uparrow)e_g^2(\uparrow)$ configuration. Moreover, the
majority-spin in $d_{z^2}$ orbital is pushed away from the Fermi
level as compared with the other orbitals, which is partly caused by
the specific crystal field. It is worth pointing that all
minority-spin are well above the Fermi level.

From Fig. 6, the majority-spin Fe-$3d$ states in $d_{xy}, d_{yz},
d_{z^2}, d_{xz}$,and $d_{x^2-y^2}$ orbitals are all occupied, and
minority-spin in corresponding orbitals are all empty. The partial
DOS of Fe-$3d$ indicates  the  high-spin(S=5/2) state of Fe$^{3+}$
with $t_{2g}^3(\downarrow)e_g^2(\downarrow)$ configuration. The
majority- and minority-spin are all pushed rather away from the
Fermi level due to applying the DFT+U, which improves the
correlation of $3d$-electrons. The minority-spin in $d_{z^2}$
orbital is far away as in Mn-$3d$ states.

From Fig.7, majority-spin for Co-$3d$ states in all five orbitals
are filled, while minority-spin in $d_{xy}, d_{xz},$ and
$d_{x^2-y^2}$ are partially occupied, and in $d_{z^2}$ are fully
filled, indicating the complicated spin configuration. Based on the
partially occupied three orbitals and fully occupied one orbitals in
minority-spin states, we regard the Co as the high-spin(S=3/2) state
of Co$^{2+}$ with
$t_{2g}^1(\downarrow)t_{2g}^3(\uparrow)e_g^1(\downarrow)e_g^2(\uparrow)$
configuration.

From Fig. 8, It can be seen the spin configuration for Ni is rather
complicated as compared with the other transition-metal elements in
the system. Majority- and minority-spin are all filled in Ni-$3d$
states. Majority-spin in $d_{x^2-y^2}$ and $d_{xz}$  are near the
Fermi level, which is partly caused by the employment of DFT+U, and
some of these majority-spin would lie above the Fermi level within
DFT. Based on the occupation of all spin states, we consider that Ni
is in low-spin(S=2/2) state of Ni$^{2+}$ with $d^8$ configuration.
Interestingly, the three majority-spin electrons have possibility to
occupy all five orbitals, and this point need to be studied further.
This feature would decrease the magnetic moment of Ni$^{2+}$, which
is in good agreement with the net magnetic moment of 1.56 $\mu_B$
calculated. Based on the assumption that the majority-spin near
Fermi level in $d_{x^2-y^2}$ and $d_{xz}$ is derived from the
conduction bands within DFT+U, we consider that the spin
configuration for Ni$^{2+}$ is
$t_{2g}^2(\uparrow)t_{2g}^3(\downarrow)e_g^1(\uparrow)e_g^2(\downarrow)$

Based on the structural analysis and spin configuration, the
toroidic moment caused by Mn and Fe is greater than that caused by
Co and Ni, which leads to a net toroidic moment and further to
ferrotoroidic characteristic, accompanying with a ferrimagnetic
configuration. This material can be used as multiferroic material
due to its magnetoelectric coupling effect, and possible candidate
for cathode material in rechargeable Li-ion batteries. For the
latter, the electronic conductivity is not only simply related to
the indirect band gap, but also related to localized small polarons
and polaron conduction.

\section{Conclusions}

Motivated by the ferrotoroidic domains observed in LiCoPO$_4$, we
predicted a new material by means of partial substitution of Co by
Mn, Fe, and Ni ions, in which the high-spin Mn and Fe ions are
placed on the sites with large toroidic radius. Therefore, this
material is expected to have excellent ferrotoroidictiy, and it can
be used as multiferroic material in terms of its magnetoelectric
effect due to the ferrotoroidicity. The new system also possesses a
ferrimagnetic configuration with magnetic moment of 1.56 $\mu_B$.
DFT+U reveal that the indirect band gap is $\sim$1.25 eV, the
electronic conductivity in Li$_4$MnFeCoNiP$_4$O$_{16}$ is not simply
related to the band gap. The valence state is found to be
Mn$^{2+}$,Fe$^{3+}$,Co$^{2+}$, and Ni$^{2+}$ for Mn, Fe, co, and Ni
ions, respectively.

 \noindent \textbf{Acknowledgement}\\
The idea for this work was inspired by a talk with Dr. Xiao-Jian
Wang.

\clearpage
\begin{table}[!h]

\caption{ Relaxed lattice parameters and wyckoff positions.}

\begin{center}

\begin{tabular}{@{}cccccccc}
\hline\hline
  $a$({\AA})    & $b$ ({\AA}) & $c$ ({\AA})  & $V$({\AA})$^3$  & \multicolumn{4}{c}{ Wyckoff positions} \\
\hline
10.3324&  6.0105 &  4.6922& 291.3992 &  Li&  0.0000  &   0.0000  &  0.0000\\
&&  &   &  Mn &  0.7787  &  0.2500  &  0.5210 \\
&&  &   &  Fe &  0.2213 &  0.7500  &  0.4790 \\
&&  &   &  Co &  0.2787  &  0.2500  &  0.9790\\
&&  &   &  Ni &  0.7213  &  0.7500  &  0.0210 \\
&&  &   &  P &  0.0964  &  0.2500  &  0.4218 \\
&&  &   & O1 &  0.0997 &  0.2500  &  0.7476 \\
&&  &   &  O2 &  0.4546 &  0.2500  &  0.2019\\
&&  &   &  O3 &  0.1704  &  0.0415  &  0.2804 \\

\hline\hline
\end{tabular}
\end{center}
\end{table}

\begin{table}[!h]

\caption{AFM magnetic configuration and calculated differences in
total energy ($\triangle E$)  in the NM, AFM1, AFM2, AFM3, and AFM4
configuration relative to its total energy in the FM configuration
in units of eV/formular unit.The four atomic sites to different AFM
configuration are Co, Mn, Ni, and Fe anticlockwise. The y axis is
set as the positive direction.}

\begin{center}

\begin{tabular}{@{}cccccccc}
\hline\hline

 &AFM1&AFM2&AFM3&AFM4&NM\\
magnetic configuration&$\uparrow\uparrow\downarrow\downarrow$&$\uparrow\downarrow\uparrow\downarrow$&$\downarrow\downarrow\uparrow\uparrow$&$\downarrow\uparrow\downarrow\uparrow$&\\

$\triangle E$&  -0.02&  -0.08 & -0.05 &  -0.07  &  1.35\\
\hline\hline
\end{tabular}
\end{center}
\end{table}

\clearpage

\raggedright \textbf{Figure captions:}

Fig.1 Crystal structure of Li$_4$MnFeCoNiP$_4$O$_{16}$ showing two
unit cells constructed by XCrySDen(Ref.\cite{13}).

Fig.2 The favorable antiferromagnetic configuration projected onto
the XY plane. The length  of the arrow indicates the strength of
magnetism

Fig. 3  Total DOS of Li$_4$MnFeCoNiP$_4$O$_{16}$  within (a)DFT+U
,and(b) DFT. (c) show the
 DOS of LiCoPO$_4$ within DFT+U.

 Fig.4  Electronic band structure of Li$_4$MnFeCoNiP$_4$O$_{16}$
 for the majority-spin states. The doted line indicates the Fermi energy.

 Fig. 5  Orbital-resolved DOS for Mn in
Li$_4$MnFeCoNiP$_4$O$_{16}$. (a), (b), (c), (d), and (e) show the
DOS for $d_{xy}, d_{yz}, d_{z^2}, d_{xz}$, and $d_{x^2-y^2}$
orbitals respectively. Majority-spin states are shown in the upper
portions and minority-spin states in the lower portions in all
panels. The Fermi level is set to zero.

Fig. 6  Orbital-resolved DOS for Fe in Li$_4$MnFeCoNiP$_4$O$_{16}$.
(a), (b), (c), (d), and (e) show the DOS for $d_{xy}, d_{yz},
d_{z^2}, d_{xz}$, and $d_{x^2-y^2}$ orbitals respectively.
Majority-spin states are shown in the upper portions and
minority-spin states in the lower portions in all panels. The Fermi
level is set to zero.

Fig. 7  Orbital-resolved DOS for Co in Li$_4$MnFeCoNiP$_4$O$_{16}$.
(a), (b), (c), (d), and (e) show the DOS for $d_{xy}, d_{yz},
d_{z^2}, d_{xz}$, and $d_{x^2-y^2}$ orbitals respectively.
Majority-spin states are shown in the upper portions and
minority-spin states in the lower portions in all panels. The Fermi
level is set to zero.

Fig. 8 Orbital-resolved DOS for Ni in Li$_4$MnFeCoNiP$_4$O$_{16}$.
(a), (b), (c), (d), and (e) show the DOS for $d_{xy}, d_{yz},
d_{z^2}, d_{xz}$, and $d_{x^2-y^2}$ orbitals respectively.
Majority-spin states are shown in the upper portions and
minority-spin states in the lower portions in all panels. The Fermi
level is set to zero.

\clearpage

\begin{figure}
\centering
\includegraphics[width=10cm]{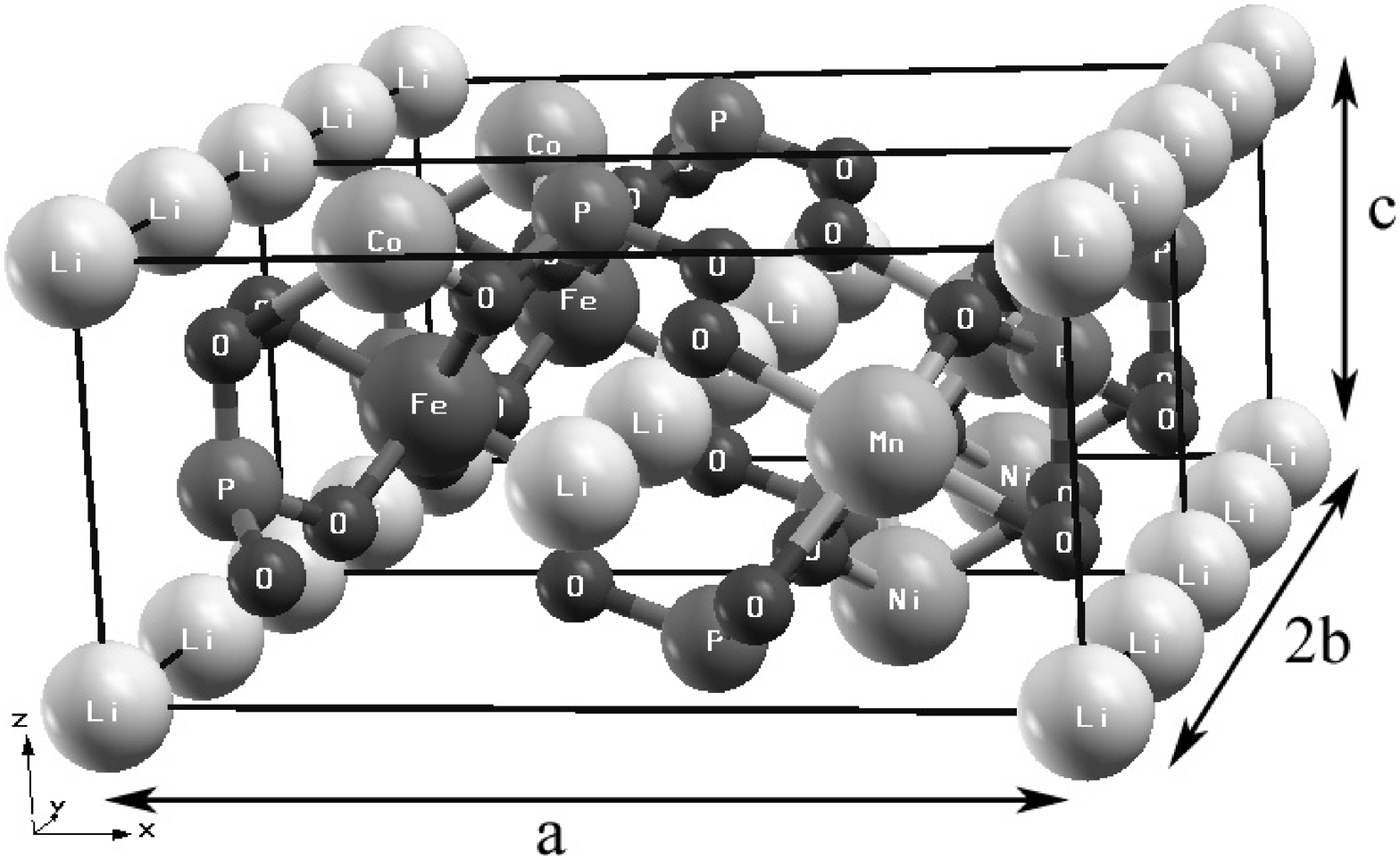}

 \caption{Crystal structure of Li$_4$MnFeCoNiP$_4$O$_{16}$ showing two unit cells constructed by XCrySDen(Ref.\cite{13}).}
\end{figure}

\begin{figure}
\centering
\includegraphics[width=10cm]{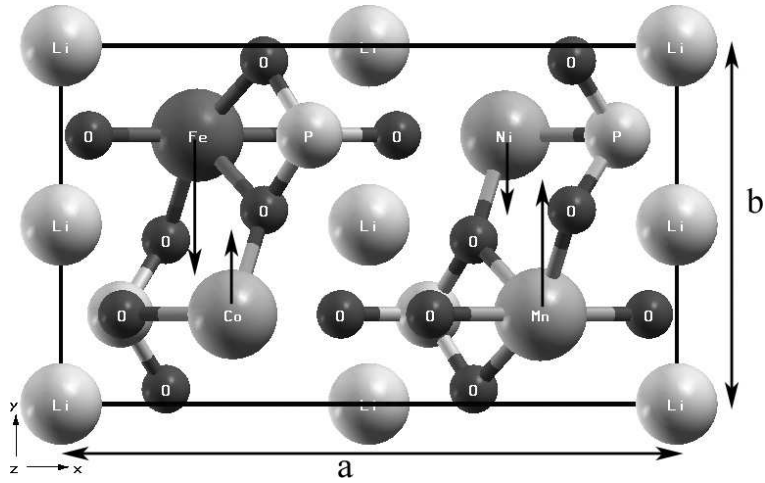}

 \caption{The favorable antiferromagnetic configuration projected onto the XY plane.The length  of the arrow indicates the strength of magnetism.}
\end{figure}

\begin{figure}
\centering
\includegraphics{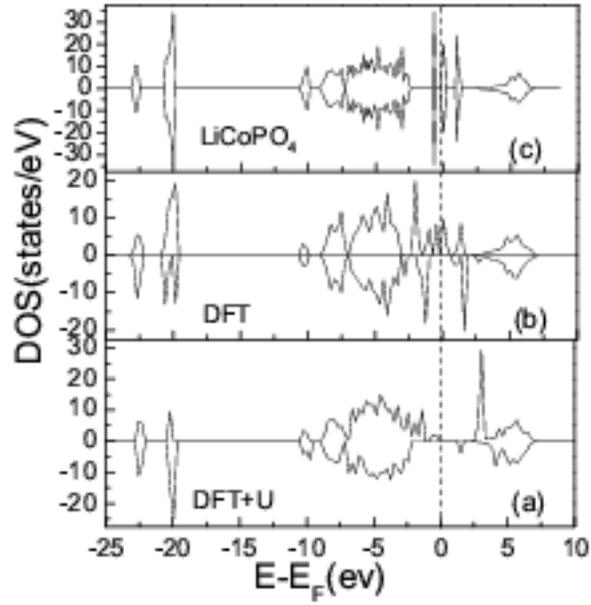}

 \caption{Total DOS of Li$_4$MnFeCoNiP$_4$O$_{16}$  within (a)DFT+U , and(b) DFT. (c) show the
 DOS of LiCoPO$_4$ within DFT+U.  }
\end{figure}

\begin{figure}
\centering
\includegraphics{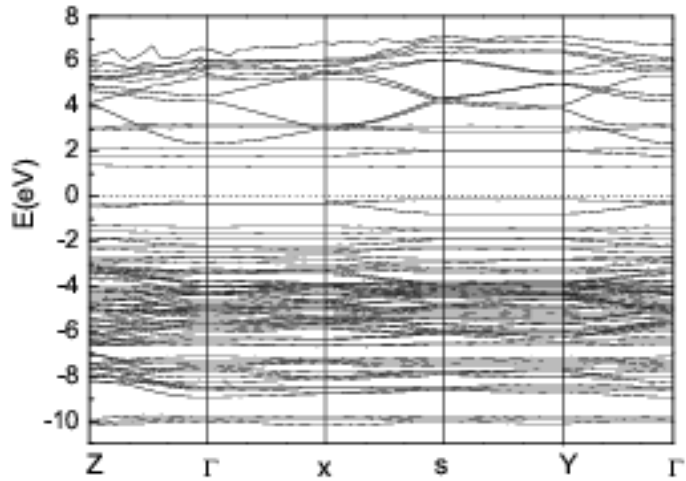}

 \caption{Electronic band structure of Li$_4$MnFeCoNiP$_4$O$_{16}$  for the majority spin states. The doted line indicates the Fermi energy.  }
\end{figure}

\begin{figure}
\centering
\includegraphics{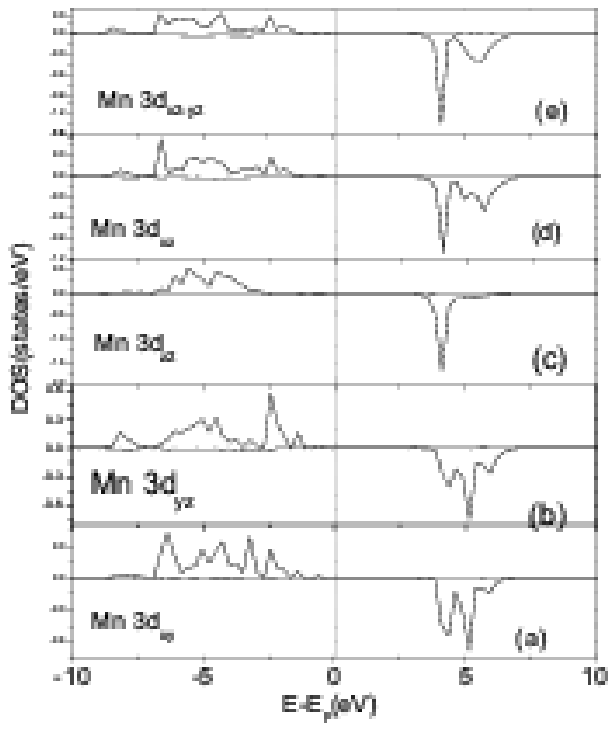}
\caption{Orbital-resolved DOS for Mn in Li$_4$MnFeCoNiP$_4$O$_{16}$.
(a), (b), (c), (d), and (e) show the DOS for $d_{xy}, d_{yz},
d_{z^2}, d_{xz}$, and $d_{x^2-y^2}$ orbitals respectively.
Majority-spin states are shown in the upper portions and
minority-spin states in the lower portions in all panels. The Fermi
level is set to zero.}
\end{figure}

\begin{figure}
\centering
\includegraphics{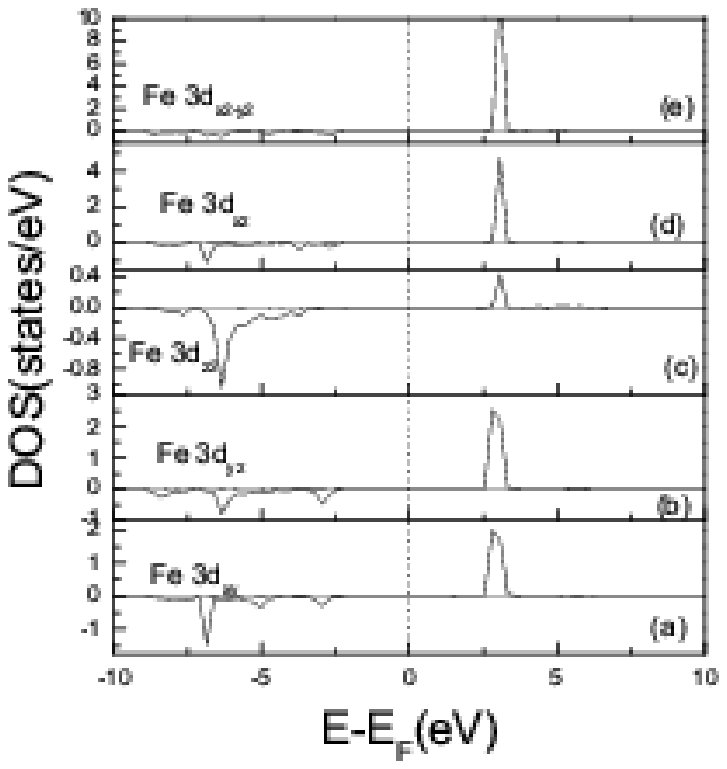}
\caption{Orbital-resolved DOS for Fe in Li$_4$MnFeCoNiP$_4$O$_{16}$.
(a), (b), (c), (d), and (e) show the DOS for $d_{xy}, d_{yz},
d_{z^2}, d_{xz}$, and $d_{x^2-y^2}$ orbitals respectively.
Majority-spin states are shown in the upper portions and
minority-spin states in the lower portions in all panels. The Fermi
level is set to zero.}
\end{figure}

\begin{figure}
\centering
\includegraphics{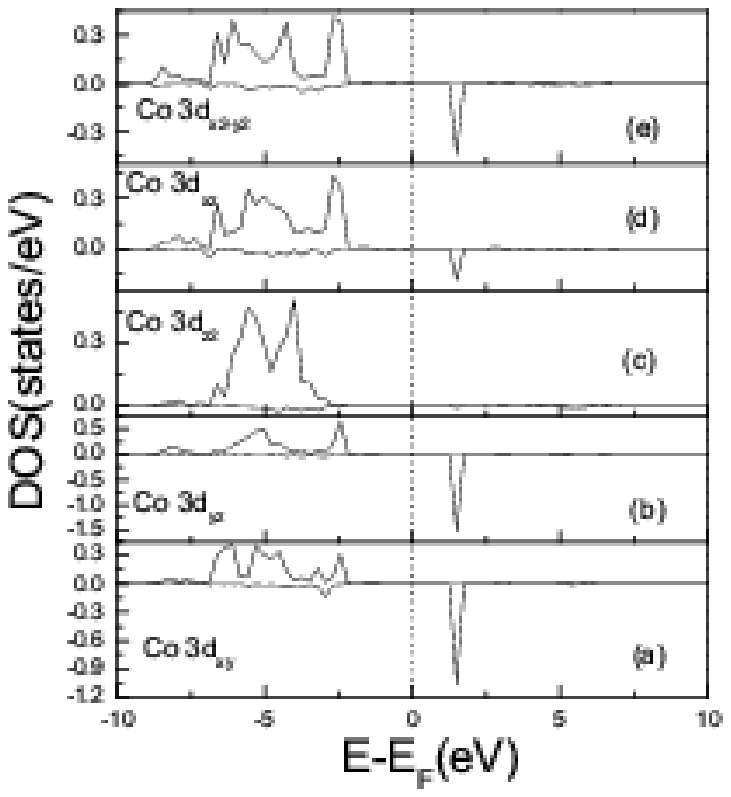}
\caption{Orbital-resolved DOS for Co in Li$_4$MnFeCoNiP$_4$O$_{16}$.
(a), (b), (c), (d), and (e) show the DOS for $d_{xy}, d_{yz},
d_{z^2}, d_{xz}$, and $d_{x^2-y^2}$ orbitals respectively.
Majority-spin states are shown in the upper portions and
minority-spin states in the lower portions in all panels. The Fermi
level is set to zero.}
\end{figure}

\begin{figure}
\centering
\includegraphics{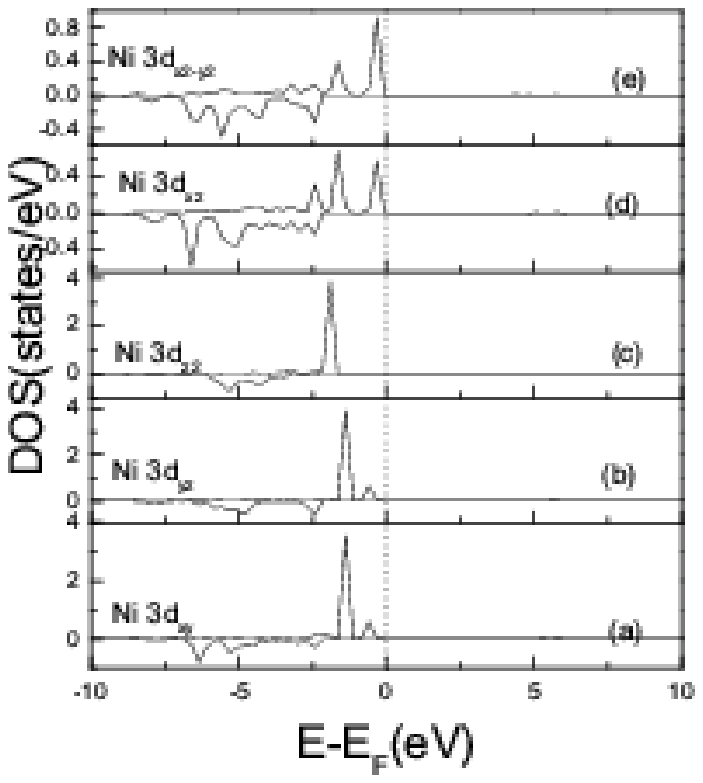}
\caption{Orbital-resolved DOS for Ni in Li$_4$MnFeCoNiP$_4$O$_{16}$.
(a), (b), (c), (d), and (e) show the DOS for $d_{xy}, d_{yz},
d_{z^2}, d_{xz}$, and $d_{x^2-y^2}$ orbitals respectively.
Majority-spin states are shown in the upper portions and
minority-spin states in the lower portions in all panels. The Fermi
level is set to zero.}
\end{figure}


\begin{thebibliography}{99}
\itemsep=-4pt
{

\bibitem{1}
B. B. Van Aken, J.-P. Rivera, H. Schmid, M. Fiebig, Nature 449
(2007) 702.

\bibitem{2}
M. Fiebig, J. Phys. D: Appl. Phys. 38 (2005) R123.


\bibitem{3}
V.M.Dubovik,V.V.Tugushev, Phys. Rep. 187 (1990) 145.



\bibitem{4}
A.A.Gorbatsevich,Y.V.Kopaev, Ferroelectrics 161 (1994) 321.





\bibitem{5}
D.G.Sannikov, Ferroelectrics 219 (1998) 177.
\bibitem{6}
H.Schmid, Ferroelectrics 252 (2001 )41.

\bibitem{7}
D.Vaknin,J.L.Zarestky,L.L.Millier,J.-P.Rivera,H.Schid, Phys. Rev. B
65 (2002) 224414.


\bibitem{8}
P. Baettig, N.A.Spaldin, Appl.Phys.Lett 86 (2005) 012505.
\bibitem{9}
H.-J. Feng , F.-M. Liu,
Phys.Lett.A(2007),doi:10.1016/j.physleta.2007.10.039.

\bibitem{10}
S.Baroni , A.Dal Corso , S.de Gironcoli , P. Giannozzi ,
C.Cavazzoni, G. Ballabio, S.Scandolo, G. Chiarotti, P.Focher, A.
Pasquarello, K.Laasonen, A.Trave, R.Car, N.Marzari,
A.Kokaljhttp://www.pwscf.org/.
\bibitem{11}
V.A.Streltsov,E.L.Belokoneva,V.G.Tsirelson,N.K.Hansen, Acta
Crystallogr., Sect.B: Struct. Sci. 49 (1993) 147.
\bibitem{12}
F.Kubel,Z.Kristallogr, Mineral. 209 (1994) 755.

\bibitem{13}
A.Kokalj, J. Mol. Graphics Model. 17 (1999) 176.

\bibitem{14}
F.Zhou,M.Cococcioni,C.A.Marianetti,D.Morgan,G.Ceder, Phys. Rev. B 70
(2004) 235121.

\bibitem{15}
T. Maxisch,F. Zhou,G. Ceder, Phys. Rev. B 73 (2006) 104301.
 }
\end{thebibliography}
\end{document}